\documentclass{aastex}
\usepackage{graphicx}
\usepackage{ulem}

\def\aap{\ {A\&A}\ }

\def\aj{\ {AJ}\ }

\def\apj{\ {ApJ}\ }
\def\apjl{\ {ApJL}\ }
\def\apjs{\ {ApJS}\ }

\def\araa{\ {ARA\&A}\ }

\def\mnras{\ {MNRAS}\ }
\def\nat{\ {Nat}\ }

\def\remove#1{{}}

\def\apgt{\ {\raise-.5ex\hbox{$\buildrel>\over\sim$}}\ }
\def\aplt{\ {\raise-.5ex\hbox{$\buildrel<\over\sim$}}\ }
\def\lteq{\ {\raise-.5ex\hbox{$\buildrel<\over-$}}\ }

\begin{document}
\title{On the minimal accuracy required for simulating
  self-gravitating systems by means of direct N-body methods}

\author{Simon Portegies Zwart}
\affil{Leiden Observatory, Leiden University, PO Box 9513, 2300 RA, Leiden,
The Netherlands}
\and

\author{Tjarda Boekholt}
\affil{Leiden Observatory, Leiden University, PO Box 9513, 2300 RA, Leiden,
The Netherlands}




\begin{abstract}
The conservation of energy, linear momentum and angular momentum are
important drivers for our physical understanding of the evolution of
the Universe. These quantities are also conserved in Newton's laws of
motion under gravity \citep{Newton:1687}. Numerical integration of the
associated equations of motion is extremely challenging, in particular
due to the steady growth of numerical errors (by round-off and
discrete time-stepping,
\cite{1981PAZh....7..752B,1993ApJ...415..715G,1993ApJ...402L..85H,1994LNP...430..131M})
and the exponential divergence
\citep{1964ApJ...140..250M,2009MNRAS.392.1051U} between two nearby
solution. As a result, numerical solutions to the general N-body
problem are intrinsically questionable
\citep{2003gmbp.book.....H,1994JAM....61..226L}.  Using brute force
integrations to arbitrary numerical precision we demonstrate
empirically that ensembles of different realizations of resonant
3-body interactions produce statistically indistinguishable
results. Although individual solutions using common integration
methods are notoriously unreliable, we conjecture that an ensemble of
approximate 3-body solutions accurately represents an ensemble of true
solutions, so long as the energy during integration is conserved to
better than 1/10.  We therefore provide an independent confirmation
that previous work on self-gravitating systems can actually be
trusted, irrespective of the intrinsic chaotic nature of the N-body
problem.
\end{abstract}

\section{Introduction}

Newton's law of gravitation is one of the fundamental laws in the
Universe that holds everything together. Although formulated in the
17th century, scientists today still study the consequences, in
particular those of many-body systems, like the solar system, star
clusters and the Milky Way galaxy.  General analytic solutions to the
N-body problem only exist for configurations
with one mass, commonly referred to as $N=1$ solutions, and for two
masses (equivalently named $N=2$, \cite{Kepler:1609, Newton:1687}).
Problems for $N\rightarrow \infty$ can be reduced via Liouville's
theorem for Hamiltonian systems to the collisionless Boltzmann
equation \citep[][but see also
  \cite{1860dMaxwell}]{1868Bolzmann,1968SvPhU..10..721V}, and
therefore analytic solutions for the global distribution function
exist.

Solutions for $N$ in between these two limits are generally realized
by computer simulations. These so-called $N$-body simulations have a
major shortcoming in that the solution to any initial realization can
only be approximated.  The main limiting factors in numerically
obtaining a true solution include errors due to round-off and
approximations both in the integration and in the time-step strategy
\citep{1996magr.meet..167K}.  These generally small errors are
magnified by the exponentially sensitive dependence on the
6N-dimensional phase-space coordinates, position and velocity
\citep{1964ApJ...140..250M}. As a consequence, the solution for a
numerically integrated self-gravitating system of $N$ masses diverges
from the true solution. This error can be controlled to some degree by
selecting a phase-space volume preserving or a symplectic algorithm
\citep{1991AJ....102.1528W} and by reducing the integration time step
\citep{1975ARA&A..13....1A,2008MNRAS.386..295H}. The latter however,
cannot be reduced indefinitely due to the accumulation of numerical
round-off in the mantissa, which is generally limited to 53 bits (64
bits in total, but 11 bits are reserved for the exponent, resulting in
only about 15 significant digits).  The exponential divergence
subsequently causes this small error to propagate to the entire system
on a dynamical time scale \citep{1993ApJ...415..715G}, which is the
time scale for a particle to cross the system once. The result of
these errors together with the exponential divergence, is the loss of
predicting power for a numerical solution to a self gravitating system
with $N>2$ after a dynamical time scale. One can subsequently question
the predicting qualities of $N$-body simulations for self-gravitating
systems, and therewith their usefulness as a scientific instrument.

We address this question for $N=3$ by brute-force numerical
integration to arbitrary precision.  The choice of $N=3$ is motivated
by the realization that this represents the first fundamental
irregular configuration with the smallest possible number of objects
that cannot be solved analytically and cannot be addressed with
collisionless theory.  In addition, 3-body
encounters form a fundamental and frequently occurring topology in any
large $N$-body simulation, and therefore also drive the global
dynamics of these larger systems.

\section{Validation of the unrestricted precision integration}
\label{Sect:validation}

The divergence between two different, approximate solutions to the
$N$-body problem can be quantified by the phase-space distance in the
positions ${\bf r}$ and velocities ${\bf v}$ of the $N$ particles (in
dimension-less N-body units):
\begin{equation}
  \delta^2 ={1 \over 6N} \sum_{i=1}^{N} \left[({\bf r}_{\rm A}-{\bf r}_{\rm B})^2 
                                      + ({\bf v}_{\rm A}-{\bf v}_{\rm B})^2\right].
\end{equation}
Values of $\delta$ are obtained by comparing the configurations from
solution $A$ and solution $B$ at any moment in time.  Each star has a
position and velocity in solution A and (generally) a different
position and velocity in solution B. For each star we calculate its
phase-space distance between the two solutions.  By dividing by $6N$,
$\delta$ can be thought of as the average difference per coordinate.
The two different runs can be performed either with the same code at a
different precision, or with two different codes, all having exactly
the same initial realization.  A value of $\delta \apgt 0.1$ indicates
that the results of the two simulations have diverged beyond
recognition.  We consider a solution to be converged to $p$ decimal
places when, for any time $t>0$, $\delta < 10^{-p}$.  (In stable
hierarchical few-body systems the value of $\delta$\, can vary
substantially across the orbital phase \citep{1986LNP...267..212D},
and one has to assure that temporary large deviations can diminish
again at a later instant.)

To investigate the build-up of numerical errors and the corresponding
exponential divergence, we developed an $N$-body solver for
self-gravitating systems which solves the N-body problem to arbitrary
precision.  This code, named {\tt Brutus} (Boekholt \& Portegies
Zwart, in preparation), is composed of a Bulirsch-Stoer integrator
\citep{springerlink:10.1007/BF01386092}, that preserves energy to the
level of the Bulirsch-Stoer tolerance. This tolerance is a parameter
which can be interpreted as the discretization error per integration
step.  The round-off error is controlled by choosing the word-length
with which all floating point numbers in the computer code are
represented. By decreasing the Bulirsch-Stoer tolerance and increasing
the word-length, we can obtain solutions to the N-body problem to
arbitrary precision.

We tested {\tt Brutus} by adopting a 3-body system of identical
particles, which are located on the vertices of an equilateral
triangle, with initial velocities such that the orbits are on a circle
around the center of mass \citep{Lagrange:1772}.  Because this system
is intrinsically unstable, small perturbations in the position and
velocity vectors cause the triangular configuration to dissolve
quickly.  The time at which this happens depends on precision.  Using
{\tt Brutus} we can reach arbitrary precision, but in this validation
experiment we stopped reducing the time step and increasing the word
length once the energy was conserved up-to 75 decimal places, which is
sufficient to demonstrate our point.  For any pre-determined time of
stability there is a combination of word-length and Bulirsch-Stoer
tolerance for which {\tt Brutus} converges.  We define a solution to
be converged when the first $p$ decimal places become independent of
the size of the time step and the word length. This is equivalent to
saying that $\delta$ is always below $10^{-p}$; for $p=3$ (at least
the first 3 digits have converged), then $\delta < 10^{-3}$ at all
times.

\section{Results}

\begin{figure}[t]
\begin{center}
\includegraphics[width=1.0\linewidth]{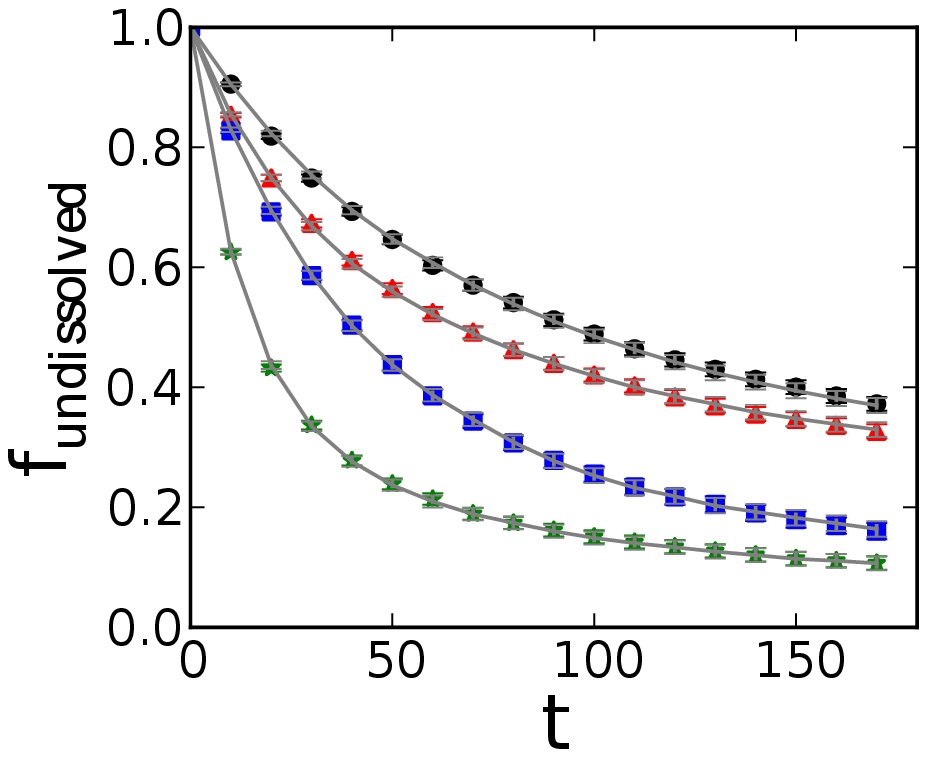}
\end{center}
\caption{Fraction of surviving systems as a function of time for the
  four sets of initial conditions.  The symbols give the results
  calculated with {\tt Brutus}, the curves give the linear
  interpolation between the points calculated with the Hermite
  integrator using $\eta = 2^{-5}$.  The virialized Plummer sphere
  with identical masses is represented by the black bullets, and with
  the range in masses as red triangles.  The blue squares and green
  stars give the results for the cold Plummer distribution without and
  with different masses, respectively.  The results of the runs with
  {\tt Hermite} are statistically indistinguishable from those with
  {\tt Brutus}. Up-to $t\simeq 50$ each of the four curves can be
  approximated with exponential decay with a half-life time of 95, 63,
  42 and 26 crossing times (from top to bottom).
\label{Fig:fdissolved}}
\end{figure}

Having established the possibility of integrating a self-gravitating
$N$-body system to arbitrary precision we can study the reliability of
N-body simulations in general.  We limit ourselves to the problem of 3
bodies, by generating a database of different 3-body problems and
solve them until a converged solution is achieved.  The positions of
the particles are taken randomly from a \cite{1911MNRAS..71..460P}
distribution and are either cold (zero kinetic energy) or
virialized. In the cold case we assured ourselves that the mutual
distances between the particles are initially comparable (within an
order of magnitude). We performed runs with identical masses and with
the masses in a ratio of 1:2:4.  For each of the four selected
ensembles of initial conditions we generated 10$^{4}$ random
realizations.  The masses and coordinates for these systems are
specified in standard double precision, to assure that the
double-precision calculations use exactly the same initial
realizations as the arbitrary-precision calculations.  Every initial
condition is integrated using the leapfrog-Verlet
(\cite{PhysRev.159.98}, we adopted the implementation available in
{\tt http://nbabel.org}) and the 4th-order Hermite predictor-corrector
scheme in a code called {\tt ph4} \citep{2012ASPC..453..129M}. (Both
codes, {\tt Brutus} and {\tt ph4} are assimilated in the public AMUSE
framework which is available at {\tt http://amusecode.org},
\cite{2013CoPhC.183..456P}).  The integration continues until the
system has been dissolved into a permanent binary and a single escaper
\citep{1975MNRAS.173..729H,1983ApJ...268..319H}.  Dissolution is
declared upon the first integral dynamical time upon which one
particle is unbound, outside a sphere of 2 initial virial radii around
the barycenter, and receding from the center of mass
\citep{1983ApJ...268..319H}. A particle is considered unbound if its
kinetic energy in the center of mass reference frame exceeds the
absolute value of its potential energy, which is more strict than
adopted in \cite{1983ApJ...268..319H}. For a fraction of the
simulations (see Fig.\,\ref{Fig:fdissolved}), the dissolution time
turns out to be very long as the evolution consists of a sequence of
ejections where a particle almost escapes, but then still returns to
once again enter a 3-body resonance.  We therefore put a constraint on
the integration time and use the fraction of long-lived systems as a
measurable statistic.  We obtain ensembles of solutions using the
Hermite and leapfrog integrators with a time-step parameter $\eta =
2^{-1}, 2^{-2}, \cdots 2^{-11}$.  Here we adopted the definition for
$\eta$ given by \cite{1975ARA&A..13....1A}.

We subsequently recalculate each of these initial realizations with
{\tt Brutus} using the same tolerance. In subsequent calculation we
systematically reduce the time-step size and increase the word-length
until we obtain a converged solution (as we discussed in
\S\,\ref{Sect:validation} for $p=3$) for every realization of the
initial conditions.  This converged solution is then compared to the
earlier simulations performed with the Hermite and leapfrog
integrators.

We now have three solutions for each initial realization of the 3-body
problem, one of which is the converged solution.  We compare the three
solutions for the time of dissolution, the semi-major axis (or
equivalently the reciprocal of the orbital energy) of the surviving
binary, its eccentricity (equivalent to the angular momentum) and the
escaper's velocity and direction.

\begin{figure}[t]
\begin{center}
\includegraphics[width=1.0\linewidth]{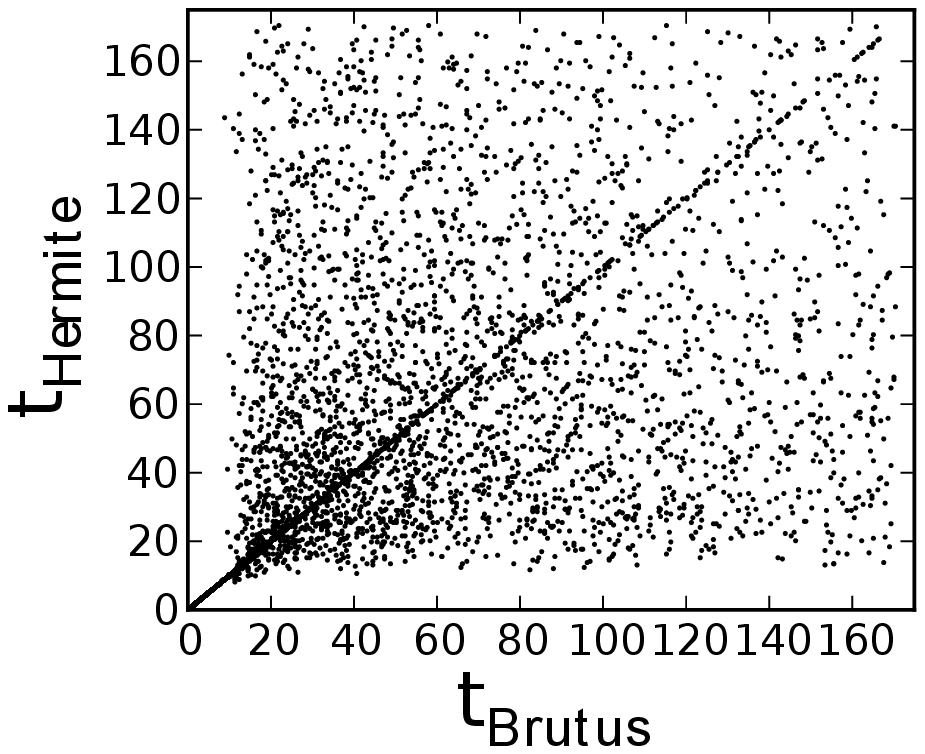}
\end{center}
\caption{Individual comparison of the dissolution time of 3-body
  systems.  Each point represents one unique initial realization of
  three equal-mass bodies taken randomly from a Plummer distribution
  in virial equilibrium.  The time to dissolution given by Hermite
  (using $\eta = 2^{-5}$) is on the ordinate and the converged value
  given by {\tt Brutus} on the abscissa. About 50$\%$ of the data
  points lie on the diagonal which represents the cases for which
  Hermite and Brutus gave very similar results. The scatter around the
  diagonal is symmetric. For very short dissolution times ($<$10
  dynamical times), there is insufficient time to grow errors and the
  results are in agreement. Once the divergence becomes important the
  Hermite integrator can return any value allowed in the experiment
  irrespective of the converged dissolution time.
\label{Fig:TimeScatter}}
\end{figure}

In Fig.\,\ref{Fig:TimeScatter} we individually compare the time to
dissolution for a certain initial realization as given by the Hermite
integrator and the converged solution as given by {\tt Brutus}. About
half of the individual Hermite solutions lie along the diagonal
representing the accurate solutions. The other half is scattered
around the diagonal. These solutions have diverged away from the
converged solution, producing a binary and an escaper with completely
different properties. For dissolutions within $\sim 10$ dynamical
times, there is insufficient time for the solution to diverge and the
results of the various numerical methods are consistent.  But once the
Hermite or leapfrog solutions have diverged away from the converged
solution the entire parameter space of the numerical experiment is
sampled.  A similar statement holds when instead of comparing the
dissolution time, we compare the properties of the binaries or the
escapers.

\begin{figure}[t]
\begin{center}
\includegraphics[width=1.0\linewidth]{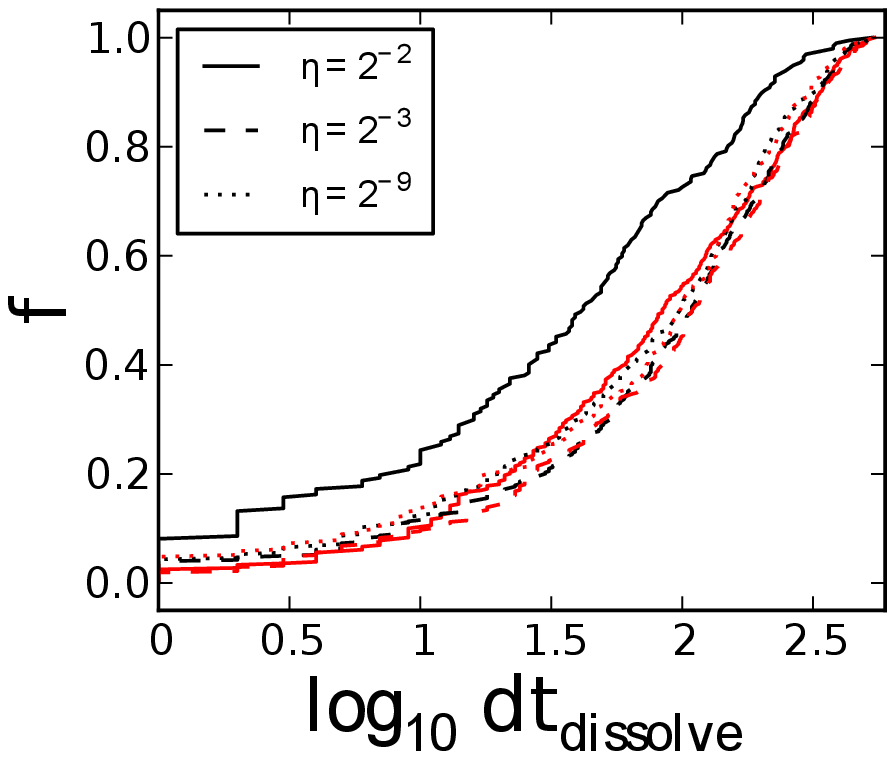}
\end{center}
\caption{Cumulative distributions for the difference in time until
  dissolution for Hermite compared to converged {\tt Brutus} solutions
  for three different values of $\eta=2^{-2}$ (solid), $\eta=2^{-3}$
  (dashes) and $\eta=2^{-9}$ (dotted curve).  The black curves give
  the distribution where the interaction calculated with Hermite
  lasted longer, and the red curves give the absolute value for the
  cases where converged solutions gave the longest lifetime for the
  interaction.  Each (black-red) pair of curves for $\eta \le 2^{-3}$
  is statistically indistinguishable, and the mean difference is
  centered around the origin.
\label{Fig:CumulativeError}}
\end{figure}

In Fig.\,\ref{Fig:CumulativeError} we present the cumulative
distribution function of the difference between the time to
dissolution of the Hermite and {\tt Brutus} calculations: $dt_{\rm
  dissolve} = t_{\rm Hermite} - t_{\rm Brutus}$ for three different
values of $\eta=2^{-2}$, $\eta=2^{-3}$ and $\eta=2^{-9}$. The
differences for $\eta \leq 2^{-3}$ are symmetric around the origin
with a dispersion of $\sim 70~N$-body time units, but for $\eta \geq
2^{-2}$ it is not symmetric. The distributions in the differences in
semi-major axis, eccentricity and the direction of the escaper (polar
and azimuthal angle with respect to the binary plane) at the time we
stop the experiment for $\eta = 2^{-3}$ down to $\eta =2^{-11}$ are
symmetric with respect to the origin. The global distributions are
statistically indistinguishable via a Kolmogorov-Smirnov test.  We
empirically determine that for a value of the time-step parameter
$\eta=2^{-3}$ the majority of the ensemble conserves energy to better
than 1/10.

In Fig.\,\ref{Fig:fdissolved} we present the fraction of undissolved
systems in time. The colored symbols give the converged solutions,
whereas the curves give the results obtained using the Hermite
integrator. The two solutions for each ensemble of initial
realizations for $\eta \leq 2^{-3}$ (as well as those obtained with
the leapfrog integrator, not shown) are statistically
indistinguishable after comparing $10^{4}$ realizations of the initial
conditions.  The distributions obtained using $\eta \geq 2^{-2}$ are
not symmetric. 

The duration of stability was studied as a function of accuracy by
\cite{2010MNRAS.407..804U} using the Sitnikov problem
\citep{1973srmd.conf.....M}. They found that the remaining time for
the system to stay bound depends on the integration accuracy.  Our
simulations did not reveal this effect, because we study systems that
dissolve on a much shorter time-scale.
 
\section{Conclusions}

The properties of the binary and the escaper of a three-body system can
be described in a statistical way. This is consistent with the
findings in previous analytic \citep{1976MNRAS.176...63M} and
numerical \citep{2004ASPC..316...45V} studies.  This behavior was
named quasi-ergodicity by \cite{1976MNRAS.176...63M}. We confirm that
this behavior remains valid also for converged 3-body solutions.

Based on the symmetry of the distribution in dissolution times (see
Fig.\,\ref{Fig:CumulativeError}), the final parameters of the binary
and escaper, as well as the consistency of the mean and median values
of the inaccurate simulations when compared to the converged solution
(see Fig.\,\ref{Fig:CumulativeError} and Fig.\,\ref{Fig:fdissolved})
we argue that global statistical distributions are preserved
irrespective of the precision of the calculation as long as energy is
preserved to better than 1/10th of the initial energy of the system.
Although we have tested only three algorithms for solving the
equations of motion we conjecture that the statistical consistency may
be preserved also for some other direct $N^2$ methods, and these may
also require that energy and angular momentum are preserved to $\leq
1/10$th.  If such direct $N$-body methods comply to the same
statistical behavior for collision-less ($N\gg 3$) systems, it will be
interesting to investigate how also other --non-$N^2$-- algorithms,
like the hierarchical tree-method \citep{1986Natur.324..446B} or
particle-mesh methods \citep{Hockney1988} behave in this respect.

In studies of self-gravitating systems which adopt the 4th order
Hermite integrator, energy and angular momentum are generally
conserved up to $\aplt 10^{-6}$ per dynamical time. Only those
simulations in which this requirement is met are often considered
reliable and suitable for scientific interpretation.  Proof for this
seemingly conservative choice has never been provided, and it is
unknown whether or not the numerical error and the exponential
divergence are not preventing certain parts in parameter space to be
accessed, or new physically inaccessible parts in parameters space
being explored.  We argued that for the resonant 3-body problem the
error made during the integration of the equations of motion poses no
problem for obtaining scientifically meaningful results so long as
energy is conserved to better than about one-tenth of the initial
total energy of the system.  In that case resonant 3-body interactions
should be treated as an ensemble average, and individual results only
contribute statistically.

By means of numerical integration until a converged solution is
obtained we find that the statistical properties of the binary and the
escaper resulting from a 3-body resonant encounter are
deterministic. This behavior is not guaranteed to propagate to larger
$N$ (see also \cite{1992MNRAS.259..505Q}); $N>3$ requires independent
testing, because these introduce more complex solutions in the form
of, for example binary-binary outcomes and hierarchical triples.  The
more extended parameter space for increasing $N$ from 3 to $N=4$ is
quite dramatic, in particular for solving the system until a converged
solution is reached.

{\bf Acknowledgements} We would like to thank Douglas Heggie for many
in-depth discussions, but also Alice Quillen, Piet Hut, Jun Makino,
Steve McMillan, Vincent Icke and Inti Pelupessy for discussions and
comments on the manuscript, as well as the anonymous referee for
careful reading and detailed comments. This work was supported by the
Netherlands Research Council NWO (grants \#643.200.503, \#639.073.803
and \#614.061.608) and by the Netherlands Research School for
Astronomy (NOVA).  Part of the numerical computations were carried out
on the Little Green Machine at Leiden University and on the LISA
cluster at SURFSara in Amsterdam.

\end{document}